\documentstyle[epsfig,graphics]{aipproc}
\input pstricks
\input ulem.sty
\input times.sty
\begin{document}
\def\A{\mbox{ASCA\ }}   \def\B{\mbox{BeppoSAX\ }}   \def\ind{\mbox{~~~}}
\def\asm{\mbox{RXTE/ASM\ }} \def\exo{\mbox{EXOSAT\ }}
\def\src{4U1700+24/HD154791 } \def\ssrc{4U1700+24 } \def\etal{{ et al. }}
\newcommand{\lx}{\mbox{L$_X$ }} \newcommand{\ergs}{\mbox{erg s$^{-1}$ }}
\newcommand{\ergcm}{\mbox{erg cm$^{-2}$ s$^{-1}$ }}
\newcommand{\mrka}{\mbox{$^{1}$}}
\newcommand{\mrkb}{\mbox{$^{2}$}}
\newcommand{\mrkc}{\mbox{$^{3}$}}
\newcommand{\mrkd}{\mbox{$^{4}$}}
\newcommand{\mrke}{\mbox{$^{5}$}}

\title{
ASCA and BeppoSAX observations
of the peculiar X--ray source 4U1700+24/HD154791}

\bigskip\medskip

\author
{D. Dal~Fiume\mrka , N. Masetti\mrka , C. Bartolini\mrkb ,
S. Del Sordo\mrkc , F. Frontera\mrkd , A. Guarnieri\mrkb ,
M. Orlandini\mrka , E. Palazzi\mrka , A. Parmar\mrke ,
A. Piccioni\mrkb , A. Santangelo\mrkc , A. Segreto\mrkc }

\address
{ \mrka Istituto TESRE/CNR, via Gobetti 101, 40129 Bologna, Italy \\
\mrkb Dipartimento di Astronomia, Universit\'a di Bologna, via
Ranzani 1, 40127 Bologna, Italy\\
\mrkc IFCAI/CNR, via U. La Malfa 153, 90146 Palermo, Italy\\
\mrkd Istituto TeSRE and Dipartimento di Fisica, Universit\'a di
Ferrara, via Paradiso 1, 44100 Ferrara, Italy\\
\mrke Space Science Department, ESA, ESTEC, Noordwjik, The
Netherlands}

\maketitle
\begin{abstract}
The X-ray source 4U1700+24/HD154791 is one of the few galactic sources
whose counterpart is an evolved M star \cite{ddf,garcia,gaudenzi}.
In X-rays the source shows extreme erratic variability and a complex and
variable spectrum. While this strongly suggests accretion onto a compact
object, no clear diagnosis of binarity was done up to now.
We report on \A and \B X--ray broad band observations of this source and on
ground optical observations from the Loiano 1.5 m telescope.
\end{abstract}
\section{Introduction}

In optical astronomy the identification of a binary system comes in most
cases from the
observation of photometric and/or radial velocity variations.
As not all X-ray binaries have known optical
counterparts, a further effective criterium in galactic X--ray
astronomy for the identification of a binary system
with an accreting compact object was often based on the observed
X--ray luminosity. For X--ray binaries harbouring a neutron star or
possibly a black hole, luminosities \lx of the order of
10$^{34}$ -- 10$^{35}$ \ergs are easily reached.
The diagnosis of the presence of a neutron star in most cases is
directly confirmed by the observation of
pulsations or thermonuclear bursts, apart from bright persistent Low
Mass X--Ray Binaries (LMXRBs).
X--ray binaries harbouring white dwarfs also show some distinctive
features. As an example in polars and intermediate polars optical and
UV observations often reveal the distinctive
signatures of the presence of a white dwarf in the system. Orbital
periods and light curves also add unambiguous and reliable evidence of
the presence of white dwarfs in this class of X--ray binaries.\\
For a number of X--ray sources the identification of a class or even the
diagnosis of binarity is rather difficult, especially when the observed
X--ray luminosity is $\leq 10^{33}$ \ergs.\\ \src belongs to this class.
The optical counterpart was identified by Garcia et al. \cite{garcia}
as a late
type giant on the basis of the positional coincidence with a HEAO1--A3
error box. The optical spectrum of this giant looks quite normal
\cite{ddf,garcia}, even if Gaudenzi and Polcaro \cite{gaudenzi}
find some interesting and variable features in its
spectrum. Variable UV line emission was detected \cite{ddf,garcia}
in different IUE pointings, showing at last some
unusual features in the emission from this otherwise normal giant. These
high excitation lines are likely linked to the same mechanism that
produces the observed X--ray emission.
In spite of various attempts, no evidence of a binary orbit was obtained
from radial velocity analysis of optical spectra.\\
The X--ray source shows extreme erratic variability, but no pulsations
were detected. The rapid (10--1000 s) time variability is strongly
suggestive of turbulent accretion, often observed in X--ray binaries.
The X--ray spectrum is rather energetic and was measured
up to 10 keV.
The X--ray luminosity \lx $\sim 10^{33}$ \ergs at an assumed
distance of 730 pc \cite{garcia} may be marginally consistent
with coronal emission, even if an evolved giant is not expected to be a
strong X-ray emitter.
Therefore the picture emerging from observations gives only hints in
favour of a binary system, given that no ``classical'' feature to be
associated to the presence of a compact object was found.\\
We have observed this source for $\sim$15 years both with X--ray satellites
(EXOSAT, \A and \B) and with ground optical observations from the Loiano
1.5 m and 0.6 m telescopes of the Bologna Astronomical Observatory.
Here we report on the \A and \B observations, performed respectively on
March 8, 1995 and on  March 27, 1998. We also report on photometric
optical UBVRI monitoring.

\section{Observations}

In Figure 1 we show the observed 1.5--9 keV count rate from the GIS2 and GIS3
instruments on board \A and the 1.5--10 keV observed count
rate from the MECS2 and MECS3 instruments on board \B.
A clear increase of the \ssrc count rate was detected in November 1997
by RXTE/ASM ({\em http://space.mit.edu/XTE/ASM\_lc.html} ).
The \B observation was performed approximately five months after this
event, when the source had already recovered its quiescent flux.
The substantial erratic variability already detected with \exo \cite{ddf}
is clearly present also in both observations. The source
flux in the \B observation is significantly lower than that in the \A
observation.
\begin{figure}[h]
\centerline{
\psfig{file=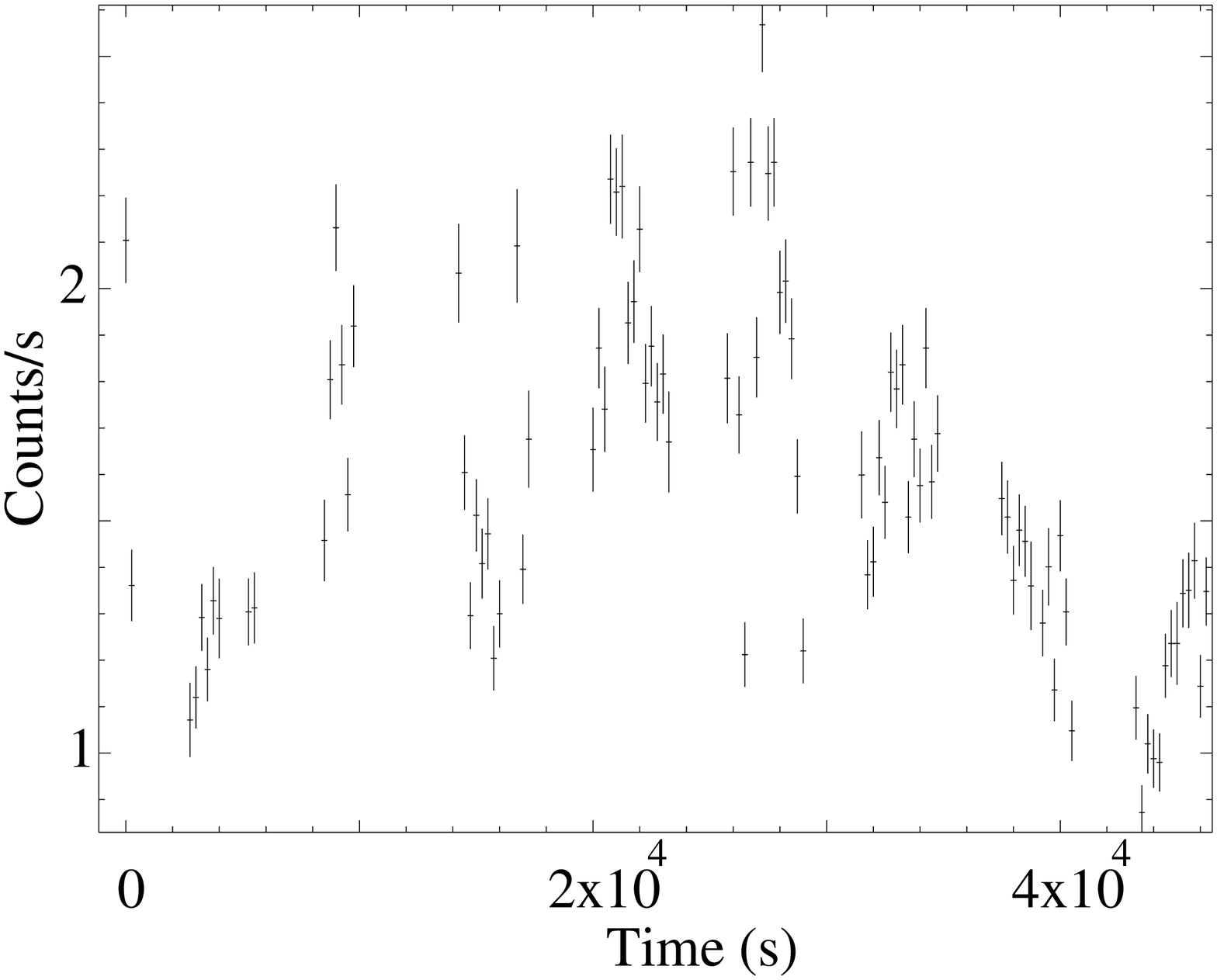,width=0.48\textwidth}\ \psfig{file=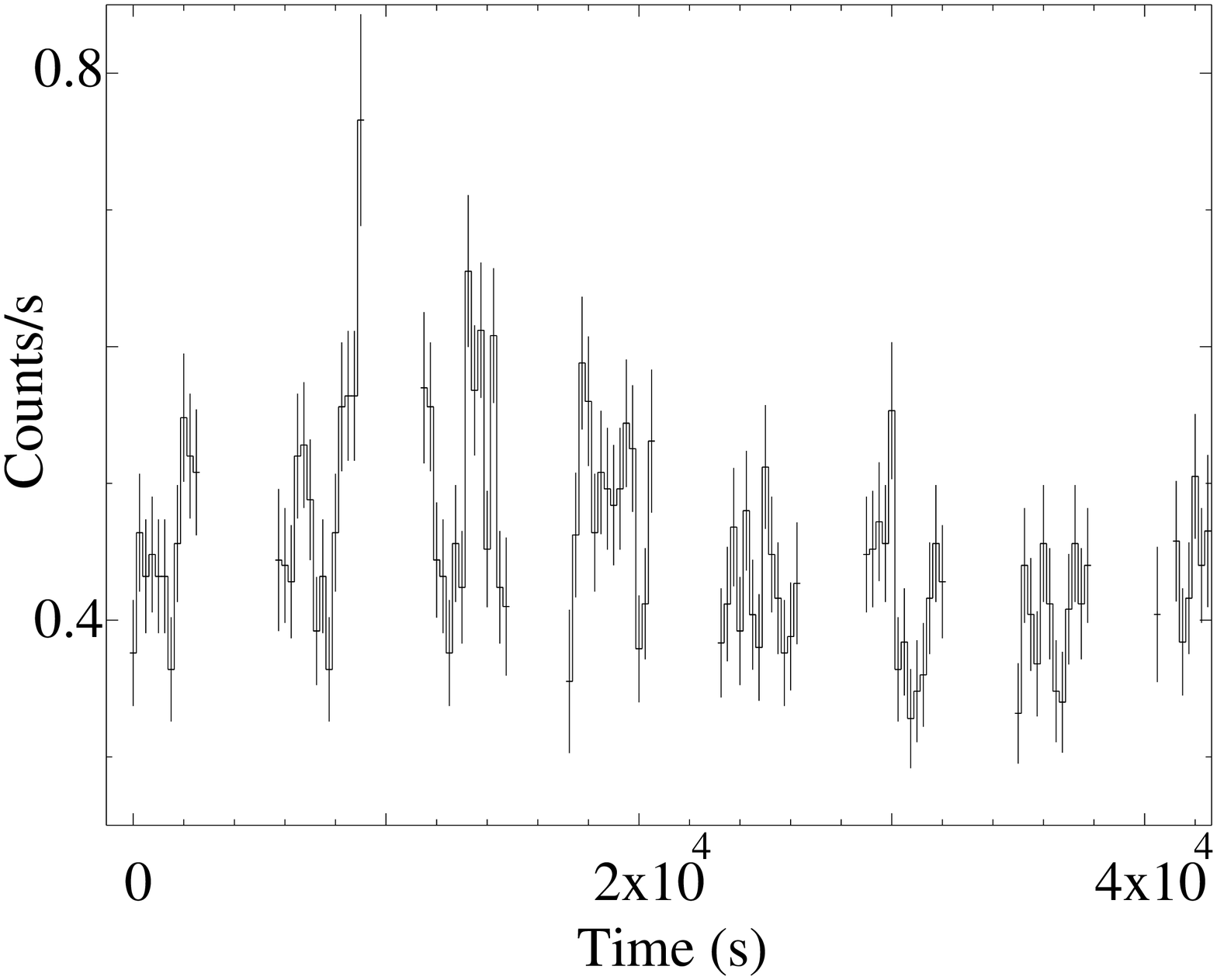,width=0.48\textwidth}
}
\caption{Count rate time series from the \A and \B observations}
\end{figure}
The erratic source variability is clearly visible in the Power Spectral
Density (PSD) shown in Figure 2, calculated on the time series of GIS2
and GIS3 count rate binned on 0.1 s. The spectra were calculated for
runs with typical duration of 3000s. The PSD shown in Figure 2 is
obtained by averaging the spectra of different runs and by summing
adjacent frequencies with a logarithmic rebinning.
\begin{figure}
\centerline{
\psfig{file=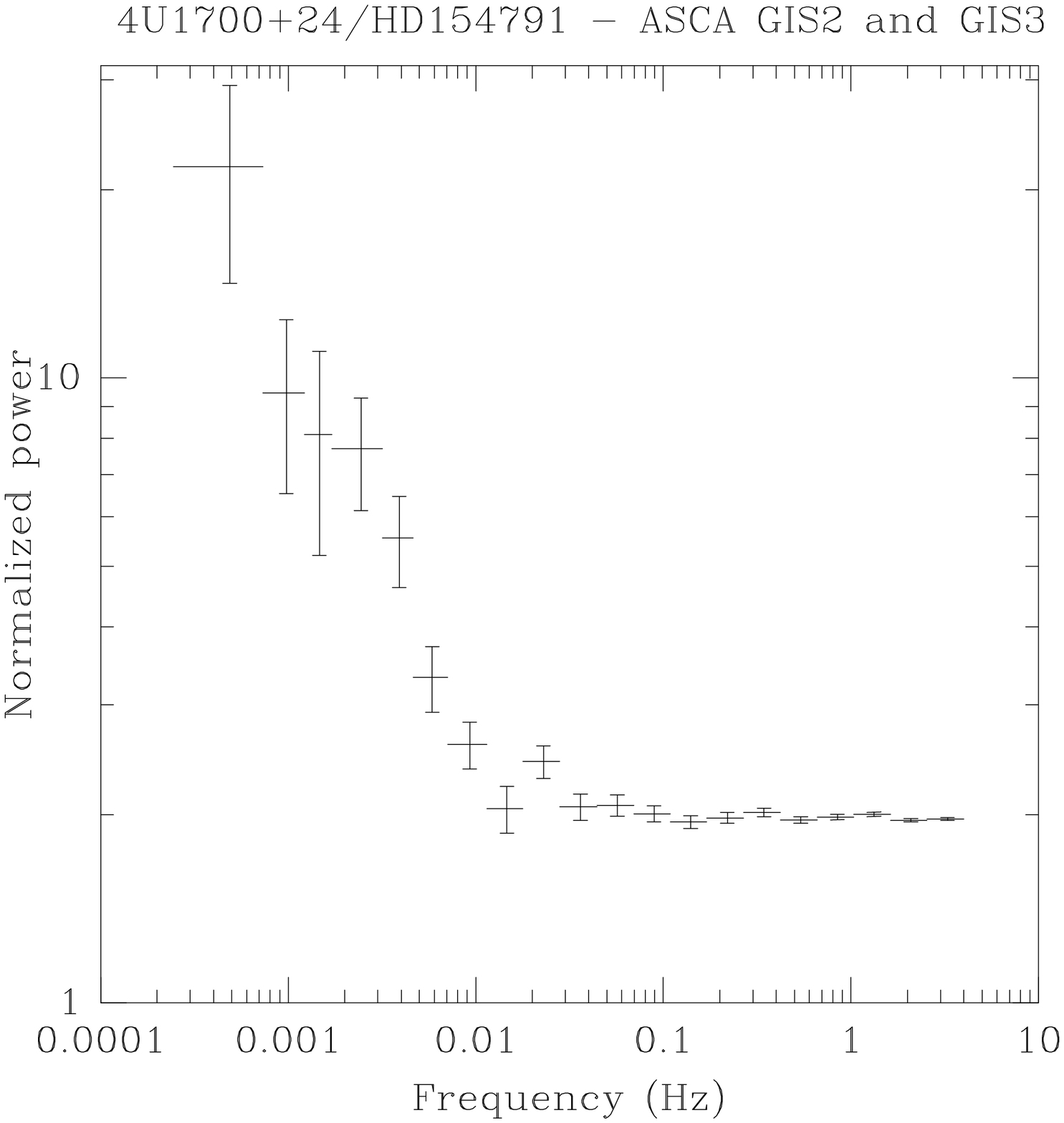,width=0.9\textwidth,height=10truecm}
}
\caption{Power Spectral Density from ASCA observation}
\end{figure}
The observed X--ray source luminosity (2--10 keV) was $L_X =
1.7\times10^{33}$ \ergs in the \A observation and $L_X=6\times 10^{32}$
\ergs in the \B observation assuming a distance of $\sim$700 pc
\cite{garcia}.\\
The X--ray energy spectrum cannot be fitted by simple single component
models. The high energy ($>$2 keV) spectrum can be fitted by an absorbed
thermal continuum, but the extrapolation of such a model at lower
energies lies significantly below the measured spectrum, both in \A
and in \B observations.\\
For a thermal model, similar to that used in the
low luminosity source $\gamma$ Cas (a suspected Be/white dwarf binary
\cite{kubo,owens}), the addition of a complex
absorber (e.g. a partial absorber) is
needed to model the low energy part of the spectrum. The lack of Fe
emission line however requires a very low Fe abundance.\\
As an example the count rate spectra from the \A and \B
observations fitted
with an optically thin thermal bremsstrahlung continuum with partial
absorber ({\it ``bremss'' } and {\it ``pcfabs'' } models in XSPEC)
are shown in Figure 3.
The \B spectrum is softer than that observed with \A, and very
similar to that observed with \exo \cite{ddf} at almost
exactly the same flux level of the \B observation.
\begin{figure}
\centerline{
\psfig{file=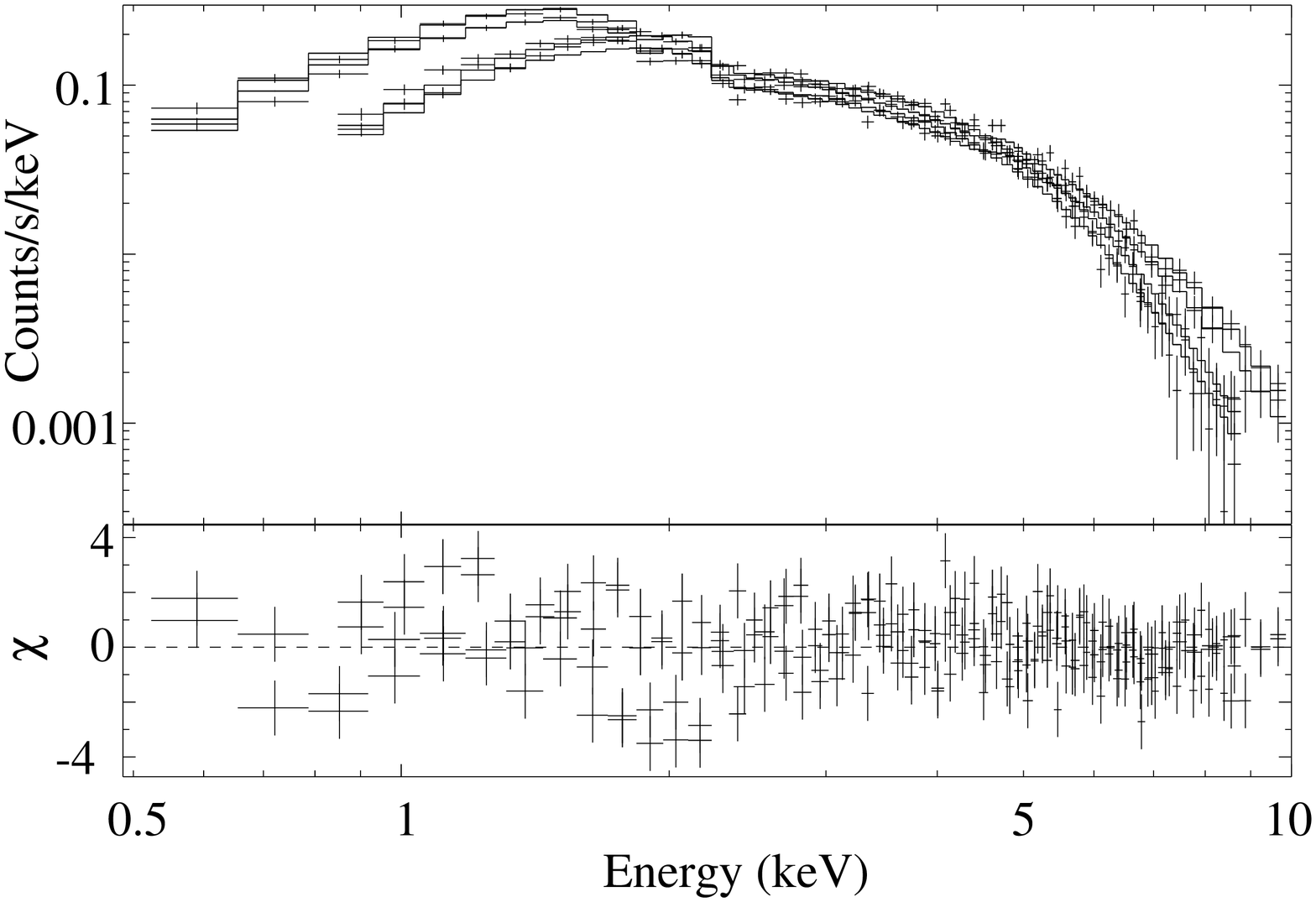,width=0.5\textwidth,height=8truecm}\ \psfig{file=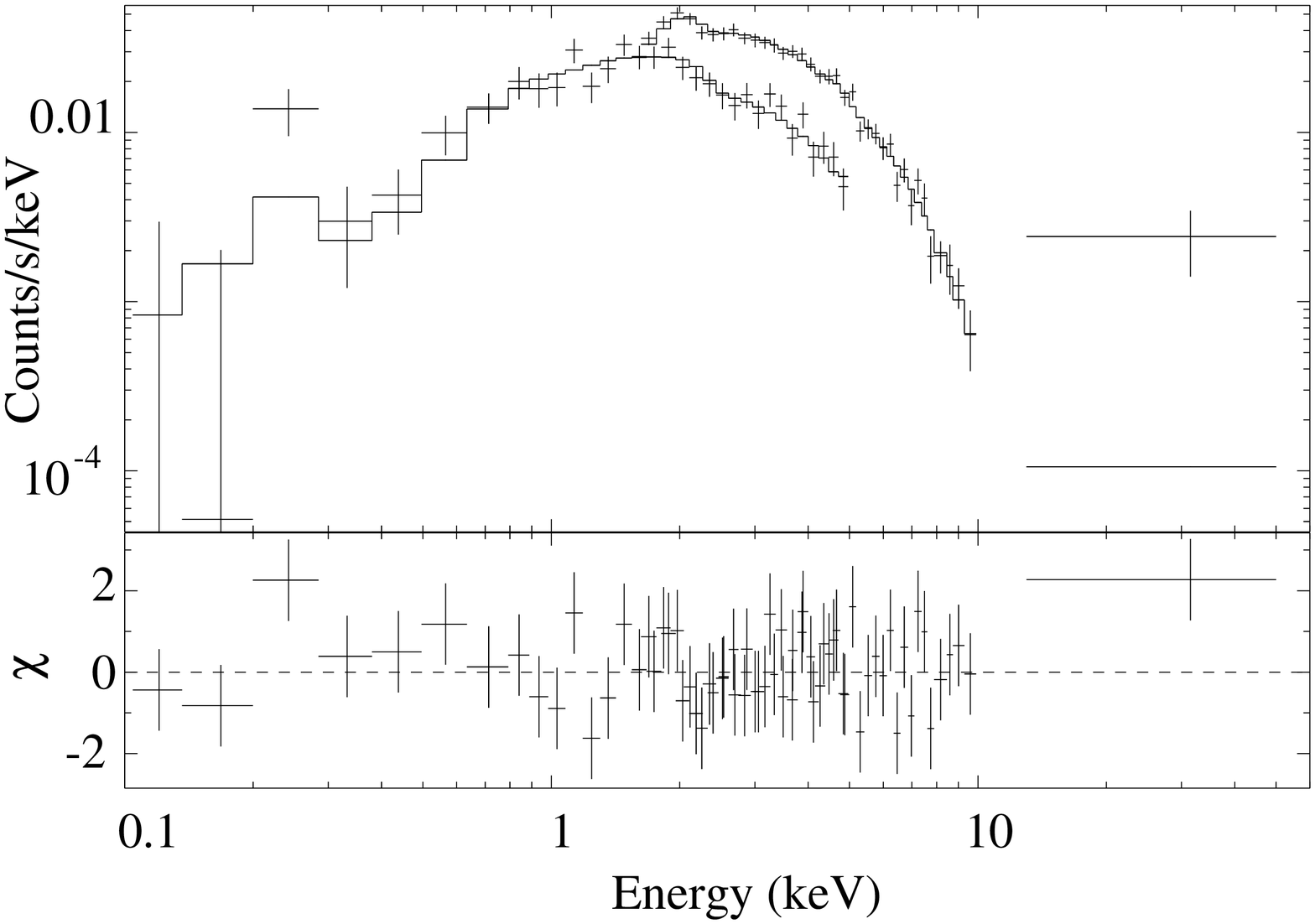,width=0.5\textwidth,height=8truecm}
}
\caption{
Left: fit to \A data. Partial covering fraction: 0.75$\pm$0.01. Temperature
kT=6.25 $\pm$ 0.2. Reduced $\chi^{2}_{dof}$: 1.55
Right: fit to \B data. Partial covering fraction: 0.72$\pm$0.04. Temperature
kT=3.6 $\pm$ 0.3. Reduced $\chi^{2}_{dof}$: 0.9
}
\end{figure}
Optical observations were performed at the Loiano 1.5 m telescope of the
Bologna Astronomical Observatory during the last 15 years.
HD154791, the optical counterpart of the X--ray source, is a M2--M3 giant
\cite{garcia,gaudenzi} with a rather normal optical
spectrum. A simple comparison with M1--M3 III templates shows
a close match with the M2 template of HD104216.
In Figure 4 we report the long term UBVRI photometry of HD154791.
No clear long term trend is visible. Some variability is present, in
particular in the U measurements, that may be intrinsic to the source.
The long-term spectral/photometric monitoring of the source is
continuing.

\begin{figure}
\centerline{\psfig{file=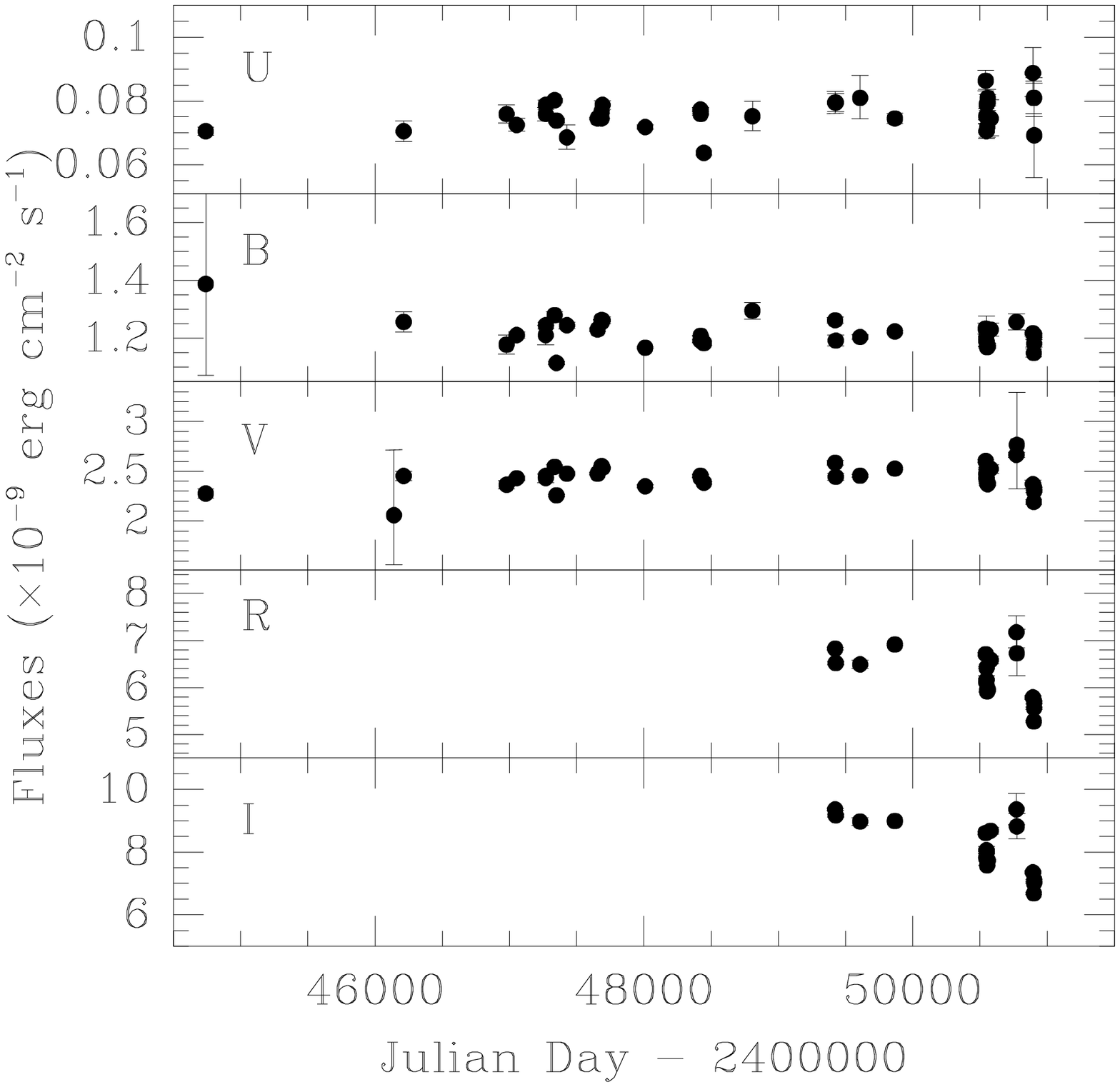,height=7.8cm,width=0.9\textwidth}}
\centerline{ }
\caption{UBVRI long term variability of HD154791}
\end{figure}

\section{Discussion}
The observations we report still cannot be used to perform a
``classical'' diagnosis of binarity. We nevertheless note some
interesting similarities with other low luminosity X--ray sources. In
particular some interesting similarities can be found with the X--ray
emission from $\gamma$ Cas.
The power spectrum is strikingly similar and the energy spectrum shows a
similar shape, even if no iron line is detected in \ssrc.

However this close resemblance of the properties of the X--ray emission
does not help to determine the presence of a compact object in a binary
system, as for $\gamma$ Cas itself the diagnosis of binarity is not
completely assessed. In fact Owens \etal \cite{owens} favour the
hypothesis of a WD binary,
but a completely different point of view is based on recent UV/X--ray
observations of $\gamma$ Cas \cite{smith}. Smith \etal support the
hypothesis that the X--ray
emission of $\gamma$ Cas comes from continuous flaring from the Be star.
This hypothesis cannot be easily adapted to the case of \src, as the
much colder M giant star should not be expected to have strong and
persistent X--ray flaring activity. If this is the case, i.e. the
observed X--ray emission from \ssrc is coronal, HD154791 should
be an exception in its own class.
If the similarity of the properties of the X--ray emission from \ssrc
and $\gamma$ Cas comes from a common origin, we suggest that the WD
binary hypothesis is much more comfortable and more easily met.

{\bf Acknowledgements}.
This research is supported by the Agenzia Spaziale Italiana (ASI) and the
Consiglio Nazionale delle Ricerche (CNR) of Italy. \B is a joint
program of ASI and of the Netherlands Agency for Aerospace Programs (NIVR).
The \A observation was performed as part of the joint ESA/Japan scientific
program. CB, AG and AP acknowledge a grant from ``Progetti di ricerca
ex-quota 60\%'' of Bologna University.

\end{document}